\documentclass[%
  preprint,
  nofootinbib,
  eqsecnum,
  superscriptaddress,
  a4paper,
  showkeys,
  showpacs,
]{revtex4-1}

\RequirePackage[T1]{fontenc}

\usepackage{amsmath}
\usepackage{amssymb}
\usepackage{hyperref}
\usepackage[dvipsnames]{xcolor}
\usepackage{xcolor}
\usepackage{slashed}
\usepackage{mathrsfs}

\begin{document}
	
	
 \title{Geometrical contribution to neutrino mass matrix}
 \author{Subhasish Chakrabarty}
\email{subhasish.chy@bose.res.in}
\author{Amitabha Lahiri}
\email{amitabha@bose.res.in}
\affiliation{S. N. Bose National Centre for Basic Sciences\\
	Block - JD, Sector - III, Salt Lake, Kolkata - 700106}


\begin{abstract}
{

The dynamics of fermions on curved spacetime requires a spin connection, which contains a part called contorsion, an auxiliary field without dynamics but fully expressible in terms of the axial current density of fermions. Its effect is the appearance of a quartic interaction of all fermions in the action, leading to a nonlinear Dirac equation involving all fermions present. Noting that left and right-chiral fermions may couple to contorsion by different strengths, we show that all fermions gain an effective mass when propagating through fermionic matter. This may have an observable effect on neutrino oscillations. In particular we find that different neutrino flavors can mix even if they have zero rest mass in vacuum, without requiring fields beyond the Standard Model.

}
\end{abstract}

\maketitle

\section{Neutrino mass and mixing}
The origin of neutrino mass is a mystery~\cite{Murayama:2006qb,King:2015aea}. The Standard Model of particle physics, so called because it is the most successful theory of all known elementary particles and the interactions between them, explains the masses of elementary particles in terms of spontaneous breaking of the $SU(2)\times U(1)$ electroweak symmetry by the vacuum expectation value (vev) of the Higgs field. 
The Higgs doublet field $\Phi = \begin{pmatrix}		\phi^+\\ \phi^0		\end{pmatrix} $ couples left-handed doublets to the right-handed singlet via the Yukawa-type interaction
%
$-h_e\left(\bar{\Psi}_{eL}\Phi e_R + \bar{e}_R\Phi^\dagger \Psi_{eL}\right)\,.$
%
For quantization, $\phi^0$ is expanded around its vev $v$ as $\phi^0 = \frac{1}{\sqrt{2}}(v + H + i\zeta)$\, with $ H\,, \zeta$ being quantum fields.
Then the Yukawa terms can be written as 
\begin{equation}\label{yukawa.e.2}
 - h_e \bigg[ \frac{v}{\surd 2} (\bar e_L e_R + \bar e_R
e_L) + \bar
\nu_{eL} e_R \phi^+ + \bar e_R \nu_{eL} \phi^-  
+ {1\over \surd 2} (\bar e e H+ i\, \bar e \gamma_5 e \zeta) 
\bigg] \,.
\end{equation}
The first term, which provides the mass of electrons, thus owes its existence to spontaneous symmetry breaking. Since the Standard Model does not include a right handed component for the neutrino, a mass term for the neutrino is not generated by these interactions. 


If neutrinos have mass, there can be mixing and oscillations between the different neutrinos, an effect that has been used to explain the solar neutrino problem, as well as the shortfall of electron-antineutrinos coming from reactors~\cite{Fukuda:1998mi, Cleveland:1998nv,Abe:2011fz,Ahmad:2001an,Ahn:2012nd,An:2012eh}. Neutrino oscillations occur because the mass eigenstates of the neutrinos are not identical with their flavour eigenstates. 
%
Let us first see what happens to neutrinos propagating in vacuum~\cite{Pontecorvo:1967fh,Gribov:1968kq,Bilenky:1978nj}. If the neutrinos are all massless and thus degenerate eigenstates of the Hamiltonian, there will be no oscillation. Suppose however that the neutrinos have mass, different masses for different species, and further that the mass eigenstates are not identical with the flavor eigenstates. 
Then there will be mixing among neutrino eigenstates, which can be parametrized by a unitary matrix~\cite{Mohapatra:1998rq}. The neutrino field $\nu_l$ which appears in a doublet with a lepton $l$ is related to the field $\nu_\alpha$ whose excitations are mass eigenstates by this matrix $U$ as\, 
%
$|\nu_{lL}\rangle = \sum_{\alpha}U_{l\alpha} |\nu_{\alpha L}\rangle\,.$
%
At time $t$\,, the flavor eigenstates are related to the mass eigenstates by
%
$|\nu_{lL}\rangle = \sum_{\alpha} e^{-iE_\alpha t} U_{l\alpha} |\nu_{\alpha L}\rangle\,.$
%
Then the probability of finding a $\nu_{l'}$ at time $t$ in a beam that had started out as $\nu_l$ is given by
\begin{equation}
P_{\nu_{l'}\nu_{l}}(t) = \left| \langle \nu_{l'}| \nu_l(t) \rangle \right|^2 
= \sum_{\alpha\,,\beta} \left| U^*_{l'\alpha} U_{l\alpha}  U^*_{l\beta} U_{l'\beta}  \right| \cos\left((E_\alpha - E_\beta)t - \phi_{ll'\alpha\beta} \right)\,,
\label{Prob-mixing}
\end{equation}
where $\phi_{ll'\alpha\beta} = \arg \left(U^*_{l'\alpha} U_{l\alpha}  U^*_{l\beta} U_{l'\beta}\right) \,.$ The neutrinos are ultrarelativistic and start with the same spatial momenta, so we can write their energies as $E_\alpha \simeq E + \frac{m_\alpha^2}{2E}$\,. We can also replace the time of travel $t$ by the distance of travel $x$ and write 
\begin{equation}\label{Prob-nu-mixing}
P_{\nu_{l'}\nu_{l}}(t) = \sum_{\alpha\,,\beta} \left| U^*_{l'\alpha} U_{l\alpha}  U^*_{l\beta} U_{l'\beta}  \right| \cos\left(\frac{( m^2_\alpha - m^2_\beta)x}{2E} - \phi_{ll'\alpha\beta} \right)\,.
\end{equation}
Clearly there will be no mixing and no oscillation if the neutrinos have vanishing mass in the vacuum. 
Interactions with a medium results in different effective masses for the neutrinos belonging to different lepton families, as first noted by Wolfenstein~\cite{Wolfenstein:1977ue}, but a neutrino mass is still needed.
On a curved spacetime however, geometry provides an additional interaction with other fermions, thus a contribution to the Hamiltonian, challenging this conclusion.

\section{Fermions in curved space}
The dynamics of fermions in curved spacetime requires a spin connection, which specifies how the covariant derivative operator acts on spinors. The Dirac matrices $\gamma^I$ are defined on an ``internal" flat space, isomorphic to the tangent space at each point. Then it is convenient to describe gravity in terms of the spin connection $A^{IJ}_\mu$ and tetrads $e^I_\mu$ in a formulation that sometimes goes by the name of Einstein-Cartan-Kibble-Sciama gravity~\cite{Cartan1922,Kibble:1961ba,Sciama:1964wt,Hehl:1976kj}. Greek indices correspond to spacetime and Latin indices belong to the internal space, the tetrads relating the metric $g_{\mu\nu}$  of spacetime with the Minkowski metric $\eta_{IJ}\, (-+++)$ of the local tangent space through the relations $\eta_{IJ}e^I_\mu e^J_\nu = g_{\mu\nu} $\,. 

In terms of these variables, the action of gravity plus (one species of) fermion can be written as 
\begin{equation}\label{action}
S = \int |e| d^4x \left[ \frac{1}{2\kappa} F^{IJ}_{\mu\nu} e^\mu_I e^\nu_J 
+ \frac{i}{2} \left( \bar{\psi}\gamma^K e^\mu_K ~D^f_\mu\psi - (\bar{\psi}\gamma^K e^\mu_K ~D^f_\mu\psi)^\dagger \right)
 +im\bar\psi\psi  \right]\,. 
\end{equation}
where $F$ is the curvature of the connection $ D \equiv d + A $\,, while the covariant derivative of a fermion has been written as $~D^f_\mu\psi = \partial_\mu \psi - \frac{i}{4} A^{IJ}_\mu \sigma_{IJ} \psi$\,. 
The cotetrad $e^\mu_I$ is defined as the inverse of the tetrad $e^I_\mu$ and satisfies $ e^\mu_I e^I_\nu = \delta^\mu_\nu$. The tetrads and the spin connection are taken to be independent fields. 
Extremizing the action with respect to the spin connection and performing some index manipulations we find an expression for the spin connection, 
\begin{equation}
A^{IJ}_\mu = \omega^{IJ}_\mu  + \frac{\kappa}{8} e^K_\mu \bar{\psi}\left[\gamma_K,\sigma^{IJ} \right]_+\psi \,. \label{spin-connection-single}
\end{equation}
Here $\omega^{IJ}_\mu $ is the part of the spin connection built purely out of tetrads.
%
%
In the metric formulation $\omega^{IJ}_\mu $ corresponds to the Levi-Civita (unique torsion-free metric-compatible) connection and is related to the Christoffel symbols as $\Gamma^\sigma_{\mu\nu} = e^{\sigma}_I\partial_\mu e^I_\nu + e^\sigma_I e_{\nu J} \omega_\mu^{IJ}\,.$ In the absence of spinorial matter the spin connection is fully described on shell by $\omega_\mu^{IJ}\,.$

If we now extremize the action with respect to the tetrads, we will get an equation into which we insert the solution for $A_\mu^{IJ}$ obtained above and contract with tetrads to produce Einstein equation
\begin{equation}\label{EE-with-spinor}
{R}_{\mu\nu} - \frac{1}{2} g_{\mu\nu}{R} = \kappa T_{\mu\nu}\,,
\end{equation}
where $R_{\mu\nu}$ and $R$ are as usual,
while the stress-energy tensor $T_{\mu\nu}$ is now quartic in the fermionic field,
\begin{eqnarray}\label{spinor-em-tensor}
\notag
T_{\mu\nu} ({\psi,\bar{\psi}}) = \frac{i}{4} && \left( \partial_{\mu} \bar{\psi} \gamma_I \psi e^I_{\nu} - \bar{\psi} \gamma_I \partial_{\mu} \psi e^I_{\nu} + \frac{i}{4} \omega^{IJ}_{\mu} e^K_{\nu} \bar{\psi}\left[\gamma_K,\sigma_{IJ}\right]_+ \psi  + (\mu\leftrightarrow\nu)  \right) \\
&&\qquad\qquad\qquad  + i m {g_{\mu\nu}} \bar{\psi}\psi - \frac{3 \kappa}{16}g_{\mu\nu} \left(\bar{\psi} \gamma^I\gamma_5\psi\right)^2\,.
\end{eqnarray}
In writing the last term we have used the identity $\left[\gamma_K,\sigma_{IJ} \right]_+  = 2\epsilon_{IJKL} \gamma^L \gamma^5\,.$
The Dirac equation in the presence of gravity is thus
\begin{equation}
 \qquad 2\gamma^K e^\mu_K  \partial_\mu \psi + e^\alpha_I \partial_\mu e^I_\alpha~\gamma^K e^\mu_K \psi + \partial_\mu e^\mu_K \gamma^K \psi 
  + 2 m\psi
- \frac{i}{4}A^{IJ}_\mu e^{\mu K} \left[\gamma_K,\sigma_{IJ} \right]_+ \psi = 0 \label{Dirac-equation-vep-1}\,.
\end{equation}

Inserting the expression for $A^{IJ}_\mu$ into this equation, we can write it as 
\begin{equation}\label{NLD2}
\gamma^K e^\mu_K  \partial_\mu \psi - \frac{i}{4}\omega^{IJ}_\mu e^{\mu K} \gamma_K \sigma_{IJ} \psi
 +  m\psi
 + \frac{3i\kappa}{8}\left(\bar{\psi}\gamma^I\gamma^5\psi\right)\gamma_I\gamma^5\psi = 0\,.
\end{equation}
This is the nonlinear Dirac equation that governs the motion of a fermion in curved spacetime. This equation has been known for a long time in various contexts for spacetimes with torsion~\cite{Gursey:1957, Finkelstein:1960, Hehl:1971qi, Gasperini:2013}. Often this equation is written in ``Planck units'' in which Planck mass and Planck length are the units of mass and length respectively, so the $\kappa$ in the nonlinear term is replaced by unity. 

However, one important point often gets overlooked or at least is not explicitly mentioned, which is the fact that every fermion field must be included in the matter action and therefore all fermions will be present in the expression for spin connection,
\begin{equation}\label{spin-connection-full}
A^{IJ}_\mu = \omega^{IJ}_\mu  + \frac{\kappa}{8} e^K_\mu\sum_f\bar{\psi_f}\left[\gamma_K,\sigma^{IJ} \right]_+\psi_f \,, 
\end{equation}
where the sum is over all species of fermions present in the universe. This term will also appear in the nonlinear Dirac equation for each type of fermion,
\begin{equation}\label{NLD3}
\gamma^K e^\mu_K  \partial_\mu \psi_i - \frac{i}{4}\omega^{IJ}_\mu e^{\mu K} \gamma_K \sigma_{IJ} \psi_i
 +  m\psi_i
 +\frac{3i\kappa}{8}\left(\sum_f\bar{\psi}_f\gamma^I\gamma^5\psi_f\right)\gamma_I\gamma^5\psi_i = 0\,.
\end{equation}

It is instructive to derive this equation from the perspective of a spacetime with torsion. If we start from a spin connection written as $A^{IJ}_\mu = \omega^{IJ}_\mu  + \Lambda^{IJ}_\mu\,,$ we find that 
\begin{equation}\label{F-torsion}
F^{IJ}_{\mu\nu}(A) = F^{IJ}_{\mu\nu}(\omega) + \partial_{[\mu}\Lambda^{IJ}_{\nu]} + \left[\omega_{[\mu}, \Lambda_{\nu]}\right] + \eta_{KL}\Lambda^{IK}_{[\mu }\Lambda^{LJ}_{\nu]} \,.
\end{equation}
This $\Lambda$ is known as contorsion, and by extremizing the action with respect to it we find that the only nonvanishing variations come from the fermionic part of the action and the last term of $F^{IJ}_{\mu\nu}(A)$, so that the equation of motion for $\Lambda$ is
\begin{equation}\label{lambda-eom}
\Lambda^{IJ}_\mu = \frac{\kappa}{8} e^K_\mu \sum_f\bar{\psi_f}\left[\gamma_K,\sigma^{IJ} \right]_+\psi_f\,.
\end{equation}
We can insert this solution for $\Lambda$ into the Einstein equations and the Dirac equation, which are then exactly the same as we have found above. Furthermore, if we substitute this expression in the action, the resulting Einstein equations and the Dirac equation are also exactly the same as found above. In general, inserting a solution into the action gives incorrect results. In this case however, the antisymmetrized covariant derivative of $\Lambda$ contribute to a total derivative in the action, so $\Lambda$ is an auxiliary field.

The action of gravity with fermions is thus
\begin{eqnarray}
S = \int |e| d^4x &&\left[\frac{1}{2\kappa}  F^{IJ}_{\mu\nu} (\omega) e^\mu_I e^\nu_J  + \frac{i}{2}\sum_f \left( \bar{\psi}_f\gamma^K e^\mu_K ~\hat{D}^f_\mu \psi_f - (\bar{\psi}_f\gamma^K e^\mu_K ~\hat{D}^f_\mu\psi_f)^\dagger  + 2 m_f \bar{\psi}_f\psi_f\right)  \right. \notag \\
&&\qquad\qquad\qquad\qquad +\left.  \frac{1}{2\kappa} \eta_{KL}\Lambda^{IK}_{[\mu }\Lambda^{LJ}_{\nu]} e^\mu_I e^\nu_J +\frac{1}{8}\sum_f e^\mu_K\Lambda_\mu^{IJ}\bar{\psi}_f\left[\gamma^K, \sigma_{IJ}\right]_+\psi_f \right]\,,\notag \\
\label{action-full}
\end{eqnarray}
where we have written $\hat{D}^f_\mu\psi = \partial_\mu\psi -\frac{i}{4}\omega^{IJ}_\mu \sigma_{IJ}\psi\,.$ 
What we have is nothing more than general relativity with fermions. The contorsion $\Lambda$ is an auxiliary field which enforces the interaction of spacetime geometry  with fermionic fields but does not propagate. In the absence of fermions $\Lambda$ vanishes, irrespective of any bosonic fields present as long as they are minimally coupled to gravity. Again this is all very well known, but writing the action in this form 
draws attention to another aspect which seems to have been overlooked. 

The invariance of this action under local Lorentz transfomations means that $\Lambda$ transforms homogeneously under them. In particular, the last term of the above action is invariant on its own. Since $\Lambda$ does not transform inhomogeneously, the coupling of $\Lambda$ to fermions is not like the coupling of a gauge field to fermions. The transformation of fermions does not affect that of $\Lambda$, so their coupling is not protected by any invariance. This way it is more analogous to the coupling of a real scalar field to fermions -- the coefficient of $\bar{\psi}\phi\psi$ can be freely set by hand.
But unlike a scalar field, $\Lambda$ can couple chirally to fermions -- it couples to the left-handed neutrinos irrespective of whether or not there are right-handed neutrinos in the universe. So there is no reason why different species of fermions cannot be coupled to $\Lambda$ with different coupling strengths, analogous to the Yukawa coupling of fermions to a scalar field.

Therfore we propose that the generic form of the action of fermions coupled to gravity must be, not~(\ref{action-full}), but
\begin{eqnarray}
S = \int |e| d^4x &&\left[\frac{1}{2\kappa} F^{IJ}_{\mu\nu} (\omega) e^\mu_I e^\nu_J  + \frac{i}{2}\sum_f \left( \bar{\psi}_f\gamma^K e^\mu_K ~\hat{D}^f_\mu \psi_f - (\bar{\psi}_f\gamma^K e^\mu_K ~\hat{D}^f_\mu\psi_f)^\dagger + 2 m_f\bar{\psi}_f\psi_f \right) \right.\notag \\
&& \left. + \frac{1}{2\kappa}\eta_{KL}\Lambda^{IK}_{[\mu }\Lambda^{LJ}_{\nu]} {e^\mu_I e^\nu_J} +\frac{1}{8}\sum_f \Lambda_\mu^{IJ} e^\mu_K\left(\lambda_{fL}\bar{\psi}_{fL} \left[\gamma^K, \sigma_{IJ}\right]_+\psi_{fL} + \lambda_{fR}\bar{\psi}_{fR} \left[\gamma^K, \sigma_{IJ}\right]_+\psi_{fR}\right) \right]\,, \notag \\
\label{action-couplings}
\end{eqnarray}
where we have taken into account the possibility that the tensor currents due to left and right-handed fermions, which transform independently under local Lorentz transformations, may couple to $\Lambda$ with different coupling constants $\lambda_{fL}$ and $\lambda_{fR}$\,, respectively. Even though in this form the action appears to be a philosophical departure from how fermions have always been treated in general relativity, it is in fact a generic form which must inevitably appear when fermions are put in curved spacetime, unless the coupling constants $\lambda_f$ are set to zero by fiat.  Furthermore, since $\Lambda$ leads to a torsion
\begin{equation}
C^\alpha_{\phantom{\alpha}\mu\nu} \equiv \Lambda^{IJ}_{[\mu} e_{\nu]J} e^\alpha_I = \frac{\kappa}{2} \epsilon^{IJKL} e^\alpha_I e_{\mu J} e_{\nu K} \sum_f \bar{\psi}_f \gamma_L \gamma_5 \psi_{f}\,, 
\end{equation}
which is totally antisymmetric and thus does not affect geodesics, all particles fall at the same rate  in a gravitational field and the principle of equivalence is not violated by these coupling constants. 

Solving for $\Lambda$ and inserting the solution back into the action as before, we get
\begin{eqnarray}
S = \int |e| d^4x &&\left[\frac{1}{2\kappa} F^{IJ}_{\mu\nu} (\omega) e^\mu_I e^\nu_J  + \frac{i}{2}\sum_f \left( \bar{\psi}_f\gamma^K e^\mu_K ~\hat{D}^f_\mu \psi_f - (\bar{\psi}_f\gamma^K e^\mu_K ~\hat{D}^f_\mu\psi_f)^\dagger  + 2 m_f\bar{\psi}_f\psi_f \right)\right.  \notag \\
&&\qquad\qquad\qquad\qquad  \left. -\frac{3\kappa}{16}\left(\sum_f \left(\lambda_{fL}\bar{\psi}_{fL} \gamma_I \gamma^5 \psi_{fL} + \lambda_{fR}\bar{\psi}_{fR} \gamma_I\gamma^5 \psi_{fR}\right)\right)^2 \right]\,.\qquad
\label{action-couplings-quartic}
\end{eqnarray}
This action of fermions in curved spacetime is the core of our proposal, we will use this as the starting point of further calculations below. It is in fact not meaningful to work with a Dirac equation containing $\Lambda$\,, because $\Lambda$ must always equal its on-shell value. Furthermore, the quartic term is independent of the background metric, but must be included as long as there is gravity in the universe. The only ways this term can be absent from the action are if gravity is turned off ($\kappa \to 0$), or if the quartic couplings $\lambda_f$ are assumed to be zero. 
This term is suppressed by two powers of Planck mass compared to the mass term, but it could still help avert gravitational  singularities~\cite{Hehl:1974cn,Gasperini:1986mv,Poplawski:2011jz,Trautman:1973wy}. We will see that it can also in principle allow neutrino oscillations even when the neutrinos are massless.

\section{Neutrino oscillations}
In considering the propagation of neutrinos through normal matter, i.e. solar or stellar cores or nuclear reactors, we need to take into account only the effects due to electrons and nucleons (or three colors each of up and down quarks) in addition to the quartic self-interaction of the neutrinos. Weak interactions will be present of course, we will come back to the effect of that. Let us also restrict to only two types of neutrinos as before. The quartic term relevant to our purpose is
\begin{eqnarray}\label{L-quartic}
{\mathscr L}_{(\bar{\psi}\psi)^2}
& = & -\frac{3\kappa}{16} \left[ \sum_{\alpha\,,\beta} \lambda_{\nu_\alpha}\lambda_{\nu_\beta} (\bar{\nu}_\alpha\gamma_I\nu_\alpha)
(\bar{\nu}_\beta\gamma^I\nu_\beta)   
\right. \notag \\ & &\qquad\qquad \left.
- 2\sum_{\alpha, f} \lambda_{\nu_\alpha}(\bar{\nu}_\alpha\gamma_I\nu_\alpha) \left(-\lambda_{fV}\bar{\psi}_{f} \gamma^I\psi_{f}
+ \lambda_{fA}\bar{\psi}_{f} \gamma^I\gamma^5\psi_{f}\right)
\right] +\cdots
\end{eqnarray}
where we have used the fact the neutrinos are left-handed, written $\lambda_V = \frac{1}{2}(\lambda_L - \lambda_R)\,, \lambda_A = \frac{1}{2}(\lambda_L + \lambda_R)\,$ for the other fermions, and indicated by dots the terms which do not involve neutrinos. It is easy to see that the $\nu_\alpha$ which appear in the above expression, i.e. those which couple to $\Lambda$ in~(\ref{action-couplings}), must be the mass eigenstates.

Following Wolfenstein~\cite{Wolfenstein:1977ue} we calculate the forward scattering amplitude of the $\alpha$-type neutrinos, 
\begin{equation}\label{forward-scattering-amplitude}
{\cal M} = -\frac{3\kappa}{8}\left( \bar{\nu}_\alpha\gamma_I\nu_\alpha  \right) \lambda_{\nu_\alpha} \left\langle\sum_{\beta}\lambda_{\nu_\beta}\bar{\nu}_\beta\gamma^I\nu_\beta + \sum_{f=e,p,n}  \left(\lambda_{fV}\bar{\psi}_{f} \gamma^I\psi_{f}
- \lambda_{fA}\bar{\psi}_{f} \gamma^I\gamma^5\psi_{f}\right) \right\rangle\,,
\end{equation}
where the average is taken over the background. In the second sum, the spatial components of the axial current average to spin in the nonrelativistic limit, which for normal matter is negligible. The axial charge is also negligible. Similarly, the spatial components of the vector current average to the spatial momentum of the background, which can also be neglected. 
Since neutrinos are ultrarelativistic, their density inside a finite volume such as a star is bounded by the rate of production times the average density of the region, i.e. several orders of magnitudes smaller than the density of electrons or baryons. Thus the average of the neutrino term can also be neglected.

So what we are left with is the average of the temporal component of the vector current of fermions, which is nothing but the number density of the 
fermions~\footnote{We are being a bit sloppy here -- the ``density" of the fermion field is the time component of $j^\mu \equiv e^\mu_I\bar\psi\gamma^I\psi$. If the spacetime allows a 3+1 decomposition of the background metric as $g_{\mu\nu} = (-\lambda^2 + h_{ij})$\,, the volume measures can be related as $\sqrt{-g} = \lambda\sqrt{h}\,,$ and $e^0_I = \lambda^{-1}\delta^0_I$\,, where $\delta^\mu_I$ is the Kronecker delta. In this case $j^0 = -\lambda^{-1}{\psi}^\dagger\psi$ which is integrated over three spatial dimensions against the volume measure $\lambda\sqrt{h}\,.$ We have assumed this decomposition.},
$\langle \bar{\psi} \gamma^0\psi\rangle = -\langle {\psi}^\dagger_{f} \psi_{f}\rangle = -n_f$\,. 
The contribution of the forward scattering amplitude to the effective Hamiltonian density is therefore
\begin{equation}\label{H_eff}
\delta {\mathscr H}_{\text{eff}} =  \left(\sum_{f=e,p,n}  \lambda_{f} n_f\,\right) \sum_\alpha \lambda_{\nu_\alpha} {\nu}^\dagger_\alpha \nu_\alpha    ,
\end{equation}
where we have now dropped the subscript $V$\, and absorbed a factor of $\sqrt{\frac{3\kappa}{8}}$ in the definition of each of the $\lambda\,.$

This term acts as an effective mass term for the neutrinos, with $m_\alpha = \lambda_{\nu_\alpha}\rho$\,, where $\rho = \sum  \lambda_{f} n_f$\, is a weighted density of fermions that is the same for all neutrinos. This effective mass term modifies the mass of the neutrino and thus the oscillation formula, but even more interestingly, this term will cause neutrino oscillations even if neutrinos are massless. In that case, with two species of neutrinos we should replace $|m^2_2 - m^2_1|$ by  $\rho^2 |\lambda^2_{\nu_2} - \lambda^2_{\nu_1}|\, $ for constant density. The mixing matrix takes the form  $U = \begin{pmatrix} \cos\theta & \sin\theta \\ -\sin\theta & \cos\theta \end{pmatrix}$\,, so the probability of conversion of one particular flavor of neutrino into the other becomes
\begin{equation}\label{Prob-conv}
P_{\mathrm{conv}} = \sin^2 2\theta \sin^2 \left(\frac{\rho^2\Delta\lambda^2}{4E} x\right)\,,
\end{equation}
where $\Delta\lambda^2 = |\lambda^2_{\nu_2} - \lambda^2_{\nu_1}|\,.$

This result is qualitatively different from the usual formula for neutrino oscillations in vacuum. If we do not write a mass term for the neutrino, all contributions to neutrino mass comes from the quartic interaction of the neutrino with fermions in the background as well as with itself. The actual background geometry of the spacetime does not contribute to the effective mass, at least for small curvatures, for which the leading order result of the forward scattering amplitude is sufficient. 
Thus a neutrino propagating through vacuum  would not oscillate into different flavors, but oscillation would occur only in the region where there is a fermion density and stop when the neutrino leaves that region.  This is exactly like what happens for oscillation due to weak interactions, except for the fact that leptons and baryons all contribute to the effective mass of neutrinos. We note that the coupling constants $\lambda$ cannot be fixed by appealing to a more fundamental theory, but are in principle measurable by looking at oscillations when the neutrinos pass through different media, such as stars with different baryon densities, or nuclear reactor cores.

A non-vanishing $\lambda_V$ for any fermion requires that the left-handed component of the fermion does not couple to torsion with the same strength as the right-handed component. Thus chiral symmetry is broken by torsion, or alternatively, by the quartic term which has its origin in spacetime geometry. We note that it is not only neutrinos, but all fermions get a contribution to their masses from this geometrical mechanism. Even if we assume that the contribution to effective mass is of the same order for all fermions in the same background matter density, the mass of very dense stars can be significantly larger than what is calculated from their baryon count. This can be expected to affect stellar models, dark matter estimates, and cosmology.
%

\section{Weak interactions}
Neutrinos passing through matter will also interact with it via electroweak gauge fields. In this case, if we look at the effective four-fermion interaction at lowest order, only the interactions with electrons are relevant. This is because the weak interaction couples flavor eigenstates of the neutrinos with other fields; $\nu_e$ couples to electrons via both charged and neutral currents, while $\nu_\mu$ couples to electrons only via the neutral current. The modification of the mixing angle due to weak interactions in normal matter is straightforward to calculate~\cite{Mohapatra:1998rq}, as we show in outline below. The effective Lagrangian due to the charged current interaction can be written as
\begin{equation}
{\mathscr L}_{\mathrm{cc}} 
 = -\frac{G_F}{\sqrt{2}} \left(\bar{\psi}_e \gamma^I (1-\gamma^5) \psi_e \right)\left(\bar{\nu}_e\gamma_I (1-\gamma^5)\nu_e\right)\,, \label{cc-eff-lag}
\end{equation}
where a Fierz identity has been used. The (elastic) forward scattering amplitude provides the contribution to the Hamiltonian, $\sqrt{2}G_F \left\langle\bar{\psi}_{e} \gamma^I (1-\gamma^5) \psi_e \right\rangle\left(\bar{\nu}_{eL}\gamma_I \nu_{eL}\right) \simeq \sqrt{2}G_F n_e \nu_{eL}^\dagger \nu_{eL}\,.$ Normal matter does not contain muons, so $\nu_\mu$ does not have a charged current interaction. 

Both flavors of neutrinos have the same neutral current interactions, so that the contribution appears as a common term to the Hamiltonian, 
\begin{equation}\label{nc-eff-V}
V_{\mathrm{nc}} = \sqrt{2}G_F \sum_{f = e, p, n} n_f\left[I^f_{3L} - 2\sin^2\theta_W Q^f\right]\,,
\end{equation}
where $I^f_{3L}$ is the third component of weak isospin for the left-handed component of the fermion $f$ and $Q_f$ is its charge. For electrically neutral normal matter, the electron and proton contributions cancel each other and we are left with only the neutron contribution, equal to $-\sqrt{2} G_F n_n/2$\, for both types of neutrinos. The Hamiltonian, diagonal in the space of mass eigenstates, can thus be written in flavor space as
\begin{equation}\label{H-for-two}
H = H_c{\mathbb I} + \frac{\Delta m^2}{4E} \begin{pmatrix} -\cos\theta & \sin\theta \\ \sin\theta & \cos\theta \end{pmatrix} + \begin{pmatrix}
\sqrt{2}G_F n_e & 0 \\ 0 & 0
\end{pmatrix}\,.
\end{equation}
Here we have written $H_c$ for the common terms in the Hamiltonian, and $\Delta m^2 = \rho^2\left|\lambda^2_{\nu_2} - \lambda^2_{\nu_1}\right|\,. $  The effective mixing angle $\tilde{\theta}$, including the effects of both the geometric and weak contributions, is thus given by 
\begin{equation}\label{eff-theta}
\tan 2\tilde{\theta} = \frac{\Delta m^2 \sin 2\theta}{\Delta m^2 \cos 2\theta - 2\sqrt{2} G_F n_e E}\,.
\end{equation}
This formula is for ultrarelativistic neutrinos, and thus valid only in regions where matter density is not too high. For regions with low matter density and $n_e \simeq n_p \geq n_n$ and $n_e \to 0\,,$ we find that the right hand side is proportional to $n_e/E$\,. For three generations of leptons we can make similar substitutions into the standard formula for neutrino oscillations. For neutrinos passing through regions where the matter density is not constant (MSW effect~\cite{Wolfenstein:1977ue, Mikheev:1986gs, Mikheev:1986wj}), nonlinearity introduces additional complications particularly for very large matter densities, since effective masses of neutrinos and thus $\Delta m^2$\,, can  vary greatly in such situations. We will not attempt to do that calculation here.

\section{Discussion}
A few remarks are in order. If we are interested only in calculating neutrino oscillations, we could take a pragmatic approach and start with Eq.~(\ref{L-quartic}) as the defining interaction term. This term is very similar to what is called non-standard neutrino interactions (NSI)~\cite{Roulet:1991sm,Bergmann:2000gp,Biggio:2009nt,Davidson:2003ha}, in this case flavor-changing in the neutrino sector. However, the geometrical origin of this interaction means that all fermions are in quartic interaction with one another, including themselves. At low energies and for matter at normal densities, the only effect of this is expected to be on neutrino dynamics as we have discussed in this paper, but at high energies as well as for high densities of matter, for example in stellar collapse or in the early universe, we can expect this interaction to play an important role.
It is also not meaningful to talk about the quartic interactions in the absence of gravity. This is related to the fact that the quartic term appears to make the model  nonrenormalizable by power counting. Because of their origin from curved spacetime, the quartic couplings contain in them a factor of $\sqrt{\kappa}$ and thus must vanish in the flat space limit. So the counterterms in curved spacetime will have to involve curvature, thus the question of renormalizability cannot be addressed without a theory of quantum gravity, as has been noted elsewhere~\cite{Mielke:2017nwt}. 

The second point is about the size of the quartic term. Is the contribution of this term to neutrino oscillations negligibly small? 
We think that this question cannot be answered purely theoretically. Unlike in the case of weak interactions, where the energy required to create $W$-boson pairs from the vacuum sets the scale of the four-fermion interaction (and the oscillation formula can be calculated directly from quantum field theory~\cite{Pal:1989xs}), here the scale is not related to the quantum dynamics of the contorsion $\Lambda$\,, which does not in fact have any dynamics. Therefore the coupling constants $\lambda$ are free and can be set only by comparison with experimental data, not from any theoretical argument. By comparing with the NSI couplings, we can expect that the $\lambda$ are one or two orders of magnitude smaller than the effective quartic couplings coming from weak interactions, i.e., than the Fermi constant. If the neutrinos are massless in vacuum, the flavor-changing interaction becomes crucial for oscillations inside matter, even if it is small.

It should be noted that the use of torsion for oscillation of massless neutrinos has been proposed earlier in~\cite{DeSabbata:1981ek}. A coupling of neutrinos to torsion analogous to the last term in Eq.~(\ref{action-full}) was proposed, with different couplings for different species of neutrinos\footnote{We thank the anonymous referee for making us aware of this work.}. In this case the torsion is proportional to the spin density of the background, which for normal matter -- i.e. if spins are not aligned -- averages to zero over macroscopic volumes, so the effect on oscillations is very small. By breaking chiral symmetry in the coupling of fermions to torsion, and by using the fact that all fermions couple to torsion, we expect to find a much larger effect. There have been proposals of nonuniversal gravitational couplings of neutrinos leading to oscillations~\cite{Gasperini:1988zf,Gasperini:1989rt}, with the nonuniversality of couplings being subject to experimental constraints. In this case the equivalence principle is violated at the quantum level. In our proposal, nonuniversality of fermion couplings is restricted to their couplings with torsion, while their coupling with background gravity is universal -- all particles continue to fall at the same rate.


\begin{acknowledgements}
It is a pleasure to thank the anonymous referees for useful suggestions and references.	A.~L. thanks P.~B.~Pal and T.~Schwetz for helpful discussions. 
\end{acknowledgements}

\end{document}